\newcommand*{\ATLASLATEXPATH}{latex/}
\documentclass[cernpreprint,texlive=2011,txfonts,UKenglish]{latex/atlasdoc}
\pdfoutput=1
\usepackage{latex/atlaspackage}
\usepackage{latex/atlasphysics}
\usepackage{\ATLASLATEXPATH atlaspackage}
\usepackage{\ATLASLATEXPATH atlasbiblatex}

\usepackage{\ATLASLATEXPATH atlascontribute}

\usepackage{\ATLASLATEXPATH atlasphysics}
\usepackage{units}
\graphicspath{{logos/}{figures/}}

\usepackage{SUSYPhotons2015-defs}

\AtlasTitle{
    \texorpdfstring{
        Search for supersymmetry in a final state containing two photons and missing transverse momentum in
        $\boldsymbol{\sqrt{s}}$~=~13~TeV $\boldsymbol{pp}$ collisions at the LHC using the ATLAS
        detector
    }{
        Search for supersymmetry in a final state containing two photons and missing transverse momentum in
        sqrt(s) = 13 TeV pp collisions at the LHC using the ATLAS
        detector
    }
}

\author{The ATLAS Collaboration}

\usepackage{authblk}
\makeatletter
\renewcommand\AB@affilsepx{, \protect\Affilfont}
\makeatother
\renewcommand\Affilfont{\itshape\small} % affiliation formatting

\date{\today}

\AtlasRefCode{SUSY-2016-04}

\PreprintIdNumber{CERN-PH-2016-146}

\AtlasJournal{European Phys. J. C}
\AtlasAbstract{
A search has been made for supersymmetry in a final state containing two photons 
and missing transverse momentum using the ATLAS detector at the Large
Hadron Collider. The search makes use of \integLumi
of proton-proton collision data collected at a
centre-of-mass energy of \unit[13]{TeV} in 2015.
Using a combination of data-driven and Monte-Carlo-based approaches,
the Standard Model background is estimated to be $0.27^{+0.22}_{-0.10}$ events.
No events are observed in the signal region;
considering the expected background and its uncertainty,
this observation implies a model-independent 95\% CL upper limit of 0.93 fb (3.0 events) on
the visible cross section due to physics beyond the
Standard Model. In the context of a generalized model of gauge-mediated
supersymmetry breaking with a bino-like next-to-lightest supersymmetric particle, this leads to a lower limit of \unit[1650]{GeV}
on the mass of a degenerate octet of gluino states, independent of the
mass of the lighter bino-like neutralino.
}

\AtlasCoverSupportingNote{Diphoton + ETmiss 2015, ATL-COM-PHYS-2015-1136}{https://cds.cern.ch/record/2051527}
\AtlasCoverTwikiURL{https://twiki.cern.ch/twiki/bin/view/AtlasProtected/SUSYEtmissPhotons2015}

\AtlasCoverCommentsDeadline{9 June 2016}

\AtlasCoverAnalysisTeam{
Leonardo Carminati,
Osamu Jinnouchi,
Stefano Manzoni,
Giovanni Marchiori,
Tobias Orthen,
Ryan Reece,
Bruce Schumm,
Giacomo Zecchinelli
}

\AtlasCoverEdBoardMember{Gregor Herten~(chair)}
\AtlasCoverEdBoardMember{Benjamin Eric Kaplan}
\AtlasCoverEdBoardMember{Michel Lefebvre}
\AtlasCoverEgroupEditors{atlas-conf-susy-2015-11-editors@cern.ch}

\AtlasCoverEgroupEdBoard{atlas-conf-susy-2015-11-editorial-board@cern.ch}

\hypersetup{pdftitle={ATLAS document},pdfauthor={The ATLAS Collaboration}}

\begin{document}

\maketitle

\tableofcontents

%\PrintAtlasContribute{0.30}
%\clearpage

%-------------------------------------------------------------------------------
\section{Introduction}
\label{sec:intro}
%-------------------------------------------------------------------------------

This paper presents a search for signatures of supersymmetry in events
containing two energetic isolated photons and large missing transverse momentum
(with magnitude denoted \MET) in \integLumi of
proton-proton ($pp$) collision data at $\sqrt{s}=13$ TeV
recorded with the ATLAS detector at the Large Hadron Collider (LHC)
in 2015. 
The results are interpreted in the context 
of general gauge mediation (GGM)~\cite{Cheung:2007es,Meade:2008wd}
models that include the production of supersymmetric partners of
Standard Model (SM) particles that possess color charge.
In all models of GGM, the lightest supersymmetric particle (LSP)
is the gravitino \gravitino (the partner of the hypothetical quantum of the
gravitational field), with a mass significantly less than \unit[1]{GeV}.
In the GGM model considered here, the decay of the supersymmetric states produced in
$pp$ collisions would proceed through the next-to-lightest supersymmetric particle
(NLSP), which would then decay to the \gravitino LSP and one or more SM particles,
with a high probability of decay into $\gamma$ + \gravitino.
All accessible supersymmetric states with the exception of the \gravitino
are assumed to be short-lived, leading to prompt production of SM particles that would be observed in the
ATLAS detector. These results extend those of prior studies with 8 TeV collision data from Run 1
by the ATLAS~\cite{ref:photon_X_2012} and CMS~\cite{ref:CMS_2012} experiments.

%-------------------------------------------------------------------------------
%\section{Gauge-mediated supersymmetry}
\label{sec:ggm}
%-------------------------------------------------------------------------------
Supersymmetry~(SUSY)~\cite{Golfand:1971iw,Volkov:1973ix,Wess:1974tw,Wess:1974jb,Ferrara:1974pu,Salam:1974ig}
introduces a symmetry between fermions and bosons, resulting in a SUSY
particle (sparticle) with identical quantum numbers, with the exception of a difference
of half a unit of spin relative to its corresponding SM partner. 
If SUSY were an exact symmetry of nature,
each sparticle would have a mass equal to that of its SM partner.
Since no sparticles have yet been observed, SUSY would have to be a broken symmetry.
Assuming $R$-parity
conservation~\cite{Farrar:1978xj},
sparticles are
produced in pairs.  These would then decay through cascades involving
other sparticles until the stable, undetectable LSP is produced, leading
to a final state with significant \MET.

Experimental signatures of gauge-mediated supersymmetry-breaking %(GMSB)
models~\cite{Dine:1981gu,AlvarezGaume:1981wy,Nappi:1982hm}
are largely determined by the nature of the NLSP.
For GGM, the NLSP is often formed from an admixture of any of the SUSY partners
of the electroweak gauge and Higgs boson states.
In this study, the NLSP ($\neutralino$) is assumed to be electrically neutral and
purely bino-like (the SUSY partner of the SM U(1) gauge boson). 
In this case,
the final decay in each of the two cascades in a GGM event would be predominantly
$\neutralino\to\gamma+\gravitino$, leading
to final states with $\gamma\gamma+\met$.

\begin{figure}[tp]
  \begin{center}
    \includegraphics[width=0.50\textwidth]{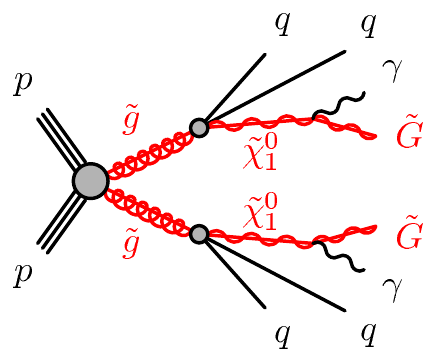} ~~
  \end{center}
  \caption{
Typical production and decay-chain processes for the gluino-pair production 
      GGM model for which the NLSP is a bino-like neutralino.
    \label{fig:feynman_bino}
  }
\end{figure}

In addition to the bino-like $\neutralino$ NLSP, a degenerate octet of gluinos (the SUSY partner
of the SM gluon) is taken to be potentially
accessible with 13 TeV $pp$ collisions. 
Both the gluino and $\neutralino$ masses are considered to be free
parameters, with the $\neutralino$ mass constrained to be less than that of the gluino.
All other SUSY masses are set to values that preclude their
production in \unit[13]{TeV} $pp$ collisions.
This results in a SUSY production
process that proceeds through the creation of pairs of 
gluino states, each of which subsequently 
decays via a virtual squark (the 12 squark flavour/chirality eigenstates
are taken to be fully degenerate) to a quark--antiquark pair plus the NLSP neutralino.
Other SM objects (jets, leptons, photons) may be produced in these cascades.
The \neutralino branching fraction
to $\gamma$ + \gravitino is 100\% for $m_{\neutralino} \to 0$ and
approaches $\cos^2 \theta_W$ for $m_{\neutralino} \gg m_Z$, with
the remainder of the \neutralino sample decaying to $Z$ + \gravitino.
For all \neutralino masses, then, the branching fraction is dominated
by the photonic decay, leading to the diphoton-plus-\met signature.
For this model with a bino-like NLSP, a typical production and decay channel for strong (gluino)
production is exhibited in Figure~\ref{fig:feynman_bino}.
Finally, it should be noted that the phenomenology relevant to this search has a negligible
dependence on the ratio $\tan\beta$ of the two SUSY Higgs-doublet vacuum expectation
values; for this analysis $\tan \beta$ is set to 1.5.

%-------------------------------------------------------------------------------
\section{Samples of simulated processes}
\label{sec:samples}
%-------------------------------------------------------------------------------

For the GGM models under study, the SUSY mass spectra and branching fractions are calculated using
\Suspect 2.41~\cite{Djouadi:2002ze} and
\Sdecay~1.3b~\cite{Muhlleitner:2003vg}, respectively, inside the package \Susyhit 1.3~\cite{Djouadi:2006bz}.
The Monte Carlo~(MC) SUSY signal samples are produced using
\Herwigpp~2.7.1~\cite{Bahr:2008pv} with {\tt CTEQ6L1} parton distribution
functions~(PDFs)~\cite{Pumplin:2002vw}.
Signal cross sections are calculated to next-to-leading order~(NLO) in the strong coupling constant,
including, for the case of strong production, the resummation of soft gluon emission at next-to-leading-logarithmic accuracy
(NLO+NLL)~\cite{Beenakker:1996ch,Kulesza:2008jb,Kulesza:2009kq,Beenakker:2009ha,Beenakker:2011fu}.
The nominal cross section and its uncertainty are taken from an envelope of cross-section predictions
using different PDF sets and factorization and renormalization scales~\cite{Botje:2011sn}.
At fixed centre-of-mass energy, SUSY production cross sections decrease rapidly with increasing SUSY particle mass.
At $\sqrt{s} = 13$ TeV, the gluino-pair production cross section is approximately \unit[25]{fb}
for a gluino mass of \unit[1.4]{TeV} and falls to below \unit[1]{fb} for a gluino
mass of \unit[2.0]{TeV}. 

While most of the backgrounds to the GGM models under examination
are estimated through the use of control samples selected from data, as described below,
the extrapolation from control regions (CRs) to the signal region (SR) depends on simulated 
samples, as do the optimization studies.
Diphoton, photon+jet, 
$W\gamma$, $Z\gamma$, $W\gamma\gamma$ and $Z\gamma\gamma$ SM processes
are generated using the {\tt SHERPA} 2.1.1 simulation package~\cite{Gleisberg:2008ta},  
making use of the {\tt CT10} PDFs~\cite{Lai:2010vv}.
The matrix elements are calculated with up to three parton emissions at leading order (four in the case of photon+jet samples)
and merged with the {\tt SHERPA} parton shower~\cite{Schumann:2007mg} 
using the ME+PS@LO prescription~\cite{Hoeche:2009rj}.
The $\ttbar\gamma$ process is generated using {\tt MadGraph5\_aMC@NLO}~\cite{Alwall:2014hca} 
with the {\tt CTEQ6L1} PDFs~\cite{Pumplin:2002vw},
in conjunction 
with {\tt PYTHIA} 8.186~\cite{Sjostrand:2007gs} with the {\tt NNPDF2.3LO} PDF set~\cite{Ball:2012cx, Ball:2013hta} 
and the {\tt A14} set~\cite{ATL-PHYS-PUB-2014-021} of tuned parameters.

All simulated samples are processed with a full ATLAS detector simulation~\cite{Aad:2010ah} based on
{\tt GEANT4}~\cite{Agostinelli:2002hh}.  The effect of additional
$pp$ interactions per bunch crossing (``pile-up'') as a function of the
instantaneous luminosity is taken into account by overlaying simulated
minimum-bias events according to the observed distribution of the
number of pile-up interactions in data, with an average of $13$
interactions per event.

\newcommand{\AtlasCoordFootnote}{%
ATLAS uses a right-handed coordinate system with its origin at the nominal interaction point (IP)
in the centre of the detector and the $z$ axis along the beam pipe.
The $x$ axis points from the IP to the centre of the LHC ring,
and the $y$ axis points upwards.
Cylindrical coordinates $(r,\phi)$ are used in the transverse plane,
$\phi$ being the azimuthal angle around the beam pipe.
The track parameter $d_0$ is defined to be the distance of closest
approach of the transverse projection of the track trajectory to the beam axis,
while the track parameter $z_0$ is defined to be the $z$ coordinate of the
track at the transverse point of closest approach.
The pseudorapidity is defined in terms of the polar angle $\theta$ as $\eta = -\ln [\tan(\theta/2)]$.
Angular distance is often specified with the quantity $\Delta R \equiv \sqrt{(\Delta\eta)^2 + (\Delta\phi)^2}$.}

%-------------------------------------------------------------------------------
\section{ATLAS detector}
\label{sec:detector}
%-------------------------------------------------------------------------------
The ATLAS experiment records $pp$ collision data with a multipurpose detector~\cite{PERF-2007-01}
that has a forward-backward symmetric cylindrical geometry and nearly
4$\pi$ solid angle coverage.\footnote{\AtlasCoordFootnote}
Closest to the beam line are solid-state tracking
devices comprising layers of silicon-based pixel and strip detectors
covering $\left|\eta\right|<2.5$ and straw-tube
detectors covering $\left|\eta\right|<2.0$, located inside a thin
superconducting solenoid that provides a \unit[2]{T} magnetic field.
Outside of this ``inner detector'', fine-grained lead/liquid-argon
electromagnetic (EM) calorimeters provide coverage over
$\left|\eta\right| < 3.2$ for the measurement of the energy and direction of
electrons and photons.
A presampler, covering $\left|\eta\right| < 1.8$, is used to correct
for energy lost upstream of the EM calorimeter.  A
steel/scintillator-tile hadronic calorimeter covers the region $|\eta|
< 1.7$, while a copper/liquid-argon medium is used for hadronic
calorimeters in the end cap region $1.5 < |\eta| < 3.2$. In the
forward region $3.2 < |\eta| < 4.9$ liquid-argon calorimeters with
copper and tungsten absorbers measure the electromagnetic and hadronic
energy.  A muon spectrometer consisting of three superconducting
toroidal magnet systems, each comprising eight toroidal coils,
tracking chambers, and detectors for
triggering, surrounds the calorimeter system.
The muon system reconstructs penetrating tracks over a range $|\eta| < 2.7$
and provides input to the trigger system over a range $|\eta| < 2.4$.
A two-level trigger system~\cite{ATL-DAQ-PUB-2016-001} is used to select events.
The first-level trigger is implemented in hardware and uses a subset of the detector information
to reduce the accepted rate to less than \unit[100]{kHz}.
This is followed by a software-based 'high-level' trigger (HLT) that
reduces the recorded event rate to approximately \unit[1]{kHz}.

%-------------------------------------------------------------------------------
\section{Event reconstruction}
\label{sec:reconstruction}
%-------------------------------------------------------------------------------

Primary vertices are formed from sets of two or more tracks,
each with transverse momentum $p_{\rm T}^{\mathrm{track}} >$ \unit[400]{MeV}, that are
mutually consistent with having originated at the same three-dimensional
point within the luminous region of the colliding proton beams.
When more than one such
primary vertex is found, the vertex with the largest sum
of the squared transverse momenta of the associated tracks is chosen.

Electron
candidates are reconstructed from EM calorimeter energy clusters
consistent with having arisen from the impact of an electromagnetic
particle (electron or photon) upon the face of the calorimeter.
For the object to be considered an electron, it is required to
match a track identified by a reconstruction algorithm
optimized for recognizing charged particles with a high probability of bremsstrahlung~\cite{ATLAS-CONF-2012-047}.
The energy of the electron candidate
is determined from the EM cluster, while its direction is
determined from the associated reconstructed track.
Electron candidates are required to have
$\pt > \unit[25]{GeV}$ and $|\eta| < 2.37$, and to be outside the transition region
$1.37 < \left|\eta\right| < 1.52$ between the central and forward portions of the
EM calorimeter.
Finally, the electron track is required to be consistent with originating from the primary vertex in both the
$r$--$z$ and $r$--$\phi$ planes.
Further details of the reconstruction of electrons can be found in
Refs.~\cite{Aad:2014fxa} and \cite{ATL-PHYS-PUB-2015-041}.

Electromagnetic clusters are classified as photon candidates provided
that they either have no matched track
or have
one or more matched tracks consistent with having arisen from a photon conversion.
Based on the
characteristics of the longitudinal and transverse shower development
in the EM calorimeter, photons are classified as
``loose'' or ``tight,'' with the tight requirements leading to a more pure
but less efficienct selection of photons relative to that of the loose requirements~\cite{Aaboud:2016yuq}.
Photon candidates are required to have $\pt > \unit[25]{GeV}$, to be within
$\left|\eta\right| < 2.37$, and to be outside the transition region
$1.37 < \left|\eta\right| < 1.52$.
Additionally, an isolation requirement is imposed:
after correcting for
contributions from pile-up and the deposition ascribed to the photon
itself, the energy within a cone of $\Delta R = 0.4$ around
the cluster barycentre is required to be less than
$\unit[2.45]{GeV} + 0.022 \times \ptg$, where
$\ptg$ is the transverse momentum of the cluster.
In the case that an EM calorimeter deposition identified as a photon 
overlaps the cluster of an identified electron within a cone of $\Delta R = 0.4$, the photon candidate is
discarded and the electron candidate is retained.
Further details of the reconstruction of photons can be found in
Ref.~\cite{Aaboud:2016yuq}.

Muon candidates make use of reconstructed tracks from the inner detector
as well as information from the muon system~\cite{ATLAS:1603.05598}.
Muons are required to be either
``combined,'' for which the muon is reconstructed independently in both the
muon spectrometer and the inner detector and then combined, or ``segment-tagged,'' for which
the muon spectrometer is used to tag tracks as muons,
without requiring a fully reconstructed candidate in the muon spectrometer.
Muons are required to have $\pt > \unit[25]{GeV}$
and $|\eta|<2.7$, with the muon track
required to be consistent with originating from the primary vertex in both the
$r$--$z$ and $r$--$\phi$ planes.

Jets are reconstructed from three-dimensional energy clusters~\cite{clustering} in the calorimeter using the anti-$k_t$
algorithm~\cite{Cacciari:2008gp} with a radius parameter $R$ = 0.4. 
Each topological cluster is calibrated to the
electromagnetic scale prior to jet reconstruction. The reconstructed jets are then calibrated to
particle level by the application of a jet energy scale derived from simulation and \emph{in situ} corrections
based on 8 TeV data~\cite{ATL-PHYS-PUB-2015-015, ATLAS-CONF-2015-037}.
In addition, the expected average energy contribution from pile-up clusters is subtracted using a factor 
dependent on the jet area~\cite{ATL-PHYS-PUB-2015-015}. Track-based selection requirements are applied 
to reject jets with $\pT{} < 60$ GeV 
and $|\eta| < 2.4$ that originate from pile-up interactions~\cite{ATLAS-CONF-2014-018}.
Once calibrated, jets are required to have $\pt >$ \unit[40]{GeV} and $|\eta| < 2.8$.

To resolve the ambiguity that arises when a photon is also reconstructed as a jet,
if a jet and a photon are reconstructed within an angular distance
$\Delta R = 0.4$ of one another, the photon is retained and
the jet is discarded.
If a jet and an electron are reconstructed within an angular distance
$\Delta R = 0.2$ of one another, the electron is retained and
the jet is discarded; if $0.2 < \Delta R < 0.4$
then the jet is retained and the electron is discarded.
Finally, in order to suppress the reconstruction of muons arising from showers induced
by jets, if a jet and a muon are found with $\Delta R < 0.4$ the jet is retained
and the muon is discarded.

The missing transverse momentum \ptm (and its magnitude, \met) is defined as the negative vector sum
of the \pt{} of all selected and calibrated physics objects in the event, with an extra term added to account
for soft energy in the event that is not associated with any of the selected objects. 
This ``\MET soft term'' is calculated
from inner-detector tracks with \pt above 400 MeV matched to the primary vertex 
to make it less dependent upon
pile-up contamination~\cite{ATLAS-CONF-2013-082, PERF-2011-03}.

Several additional observables are defined to help in the discrimination of SM backgrounds
from potential GGM signals. The total visible transverse energy,
\HT, is calculated as the scalar sum of
the transverse momenta of the selected photons and any
additional leptons and jets in the event.
The ``effective mass,'' \MEFF,
is defined as the scalar sum of \HT and \met.
The minimum jet--\ptm{} separation, \dphijm{},
is defined as the minimum azimuthal angle between
the missing transverse momentum vector and the two leading (highest-\pt)
jets with $\pt > 75$~GeV in the event, if they are present.
If no such jets exist, 
no requirement is placed on this observable.

%-------------------------------------------------------------------------------
\section{Event selection}
\label{sec:selection}
%-------------------------------------------------------------------------------

The data sample
is selected by a HLT trigger
requiring the presence of two loose photons, each with 
$\pt$ greater than 50 GeV.
Offline, two tight photons with $\pt > \unit[75]{GeV}$ are required.
In order to ensure that \MET{} is measured well, events are removed from the data sample if they
contain jets likely to be produced by beam backgrounds, cosmic rays
or detector noise~\cite{ATLAS-CONF-2015-029}.

\begin{figure}[tp]
    \centering
    \includegraphics[width=0.48\textwidth]{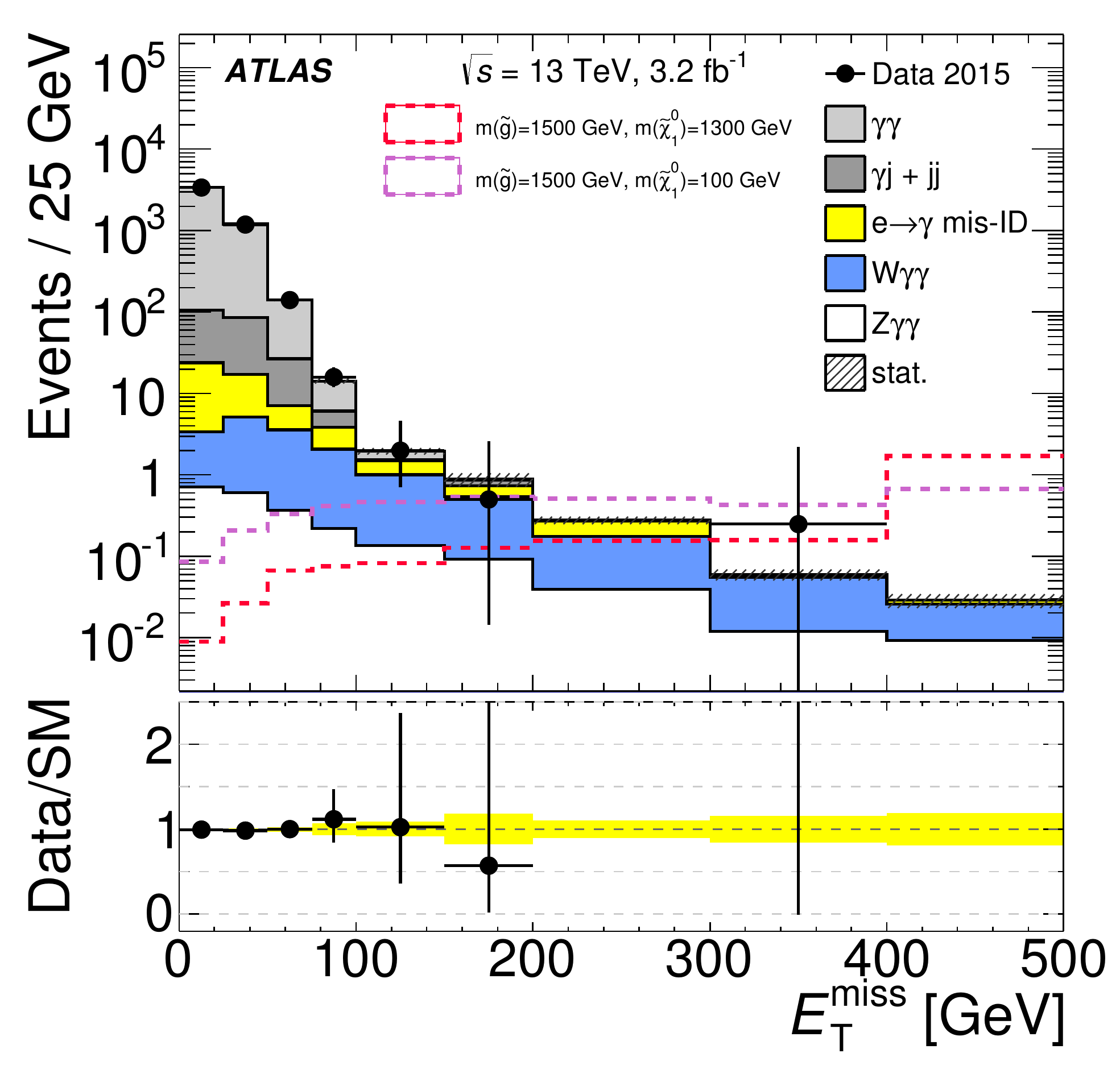} ~~
    \includegraphics[width=0.48\textwidth]{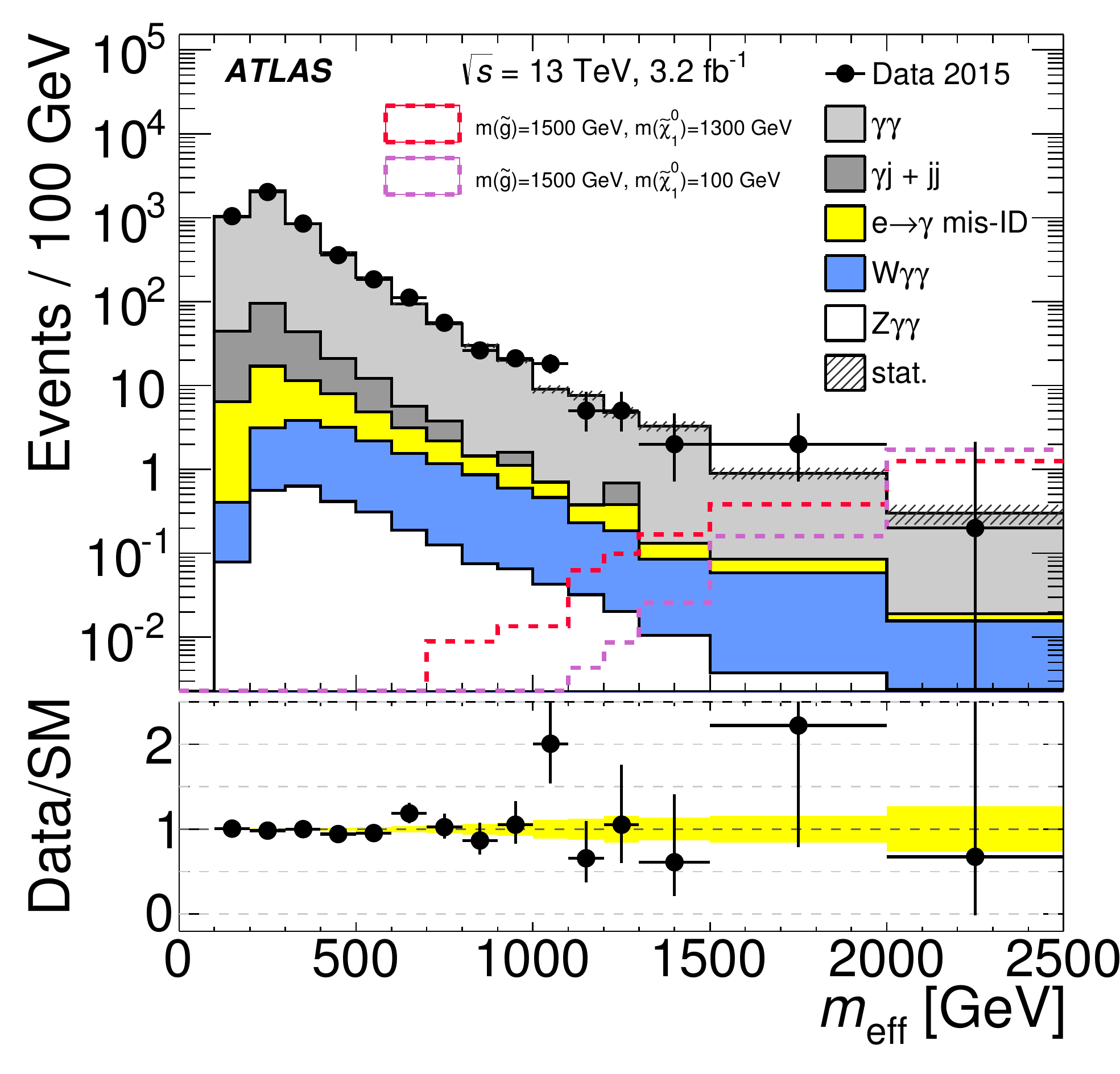} 
  \caption{
  Distributions of \MET (left) and \MEFF (right)
  for the diphoton sample after the
  application of requirements of $\ptg > 75$ GeV
  on each selected photon and of $\dphijm > 0.5$,
  but with no requirements imposed on \MET and \MEFF.
  Uncertainties (shaded
  bands for MC simulation, error bars for data) are statistical only.
  The yellow band represents the uncertainty
  in the data/SM ratio that arises from the statisical limitations of the estimates of the 
  various expected sources of SM background.
  Also shown are the
  expected contributions from the GGM signal for the two benchmark points, $(m_{\gluino},m_{\neutralino}) = (1500,1300)$ GeV and
  $(m_{\gluino},m_{\neutralino}) = (1500,100)$ GeV.
     \label{fig:met_meff}
  }
\end{figure}

To exploit the significant undetectable transverse momentum
carried away by the gravitinos, a requirement on \MET{}
is imposed on the diphoton event sample.
To take advantage of the high production energy
scale associated with signal events near the expected reach of the analysis,
an additional requirement on \MEFF{} is applied.  
To further ensure the accurate reconstruction of \MET{}
and to suppress backgrounds associated with the mismeasurement
of hadronic jets, 
a requirement of $\dphijm > 0.5$ is imposed.
Figure~\ref{fig:met_meff} shows the \MET{} and \MEFF{}
distributions of the diphoton sample after the
application of requirements of $\ptg > 75$ GeV
on each selected photon and of $\dphijm > 0.5$,
but with no requirements yet imposed on \MET{} and \MEFF{}.

As discussed in Section~\ref{sec:ggm}, the GGM signal space
is parameterized by the masses of the gluino ($m_{\gluino}$) and
bino-like NLSP ($m_{\neutralino}$). The sensitivity of this
analysis was optimized for two signal scenarios near the
expected reach in $m_{\gluino}$:
high and low neutralino-mass benchmark points were chosen with
$(m_{\gluino},m_{\neutralino}) = (1500,1300)$ GeV and
$(m_{\gluino},m_{\neutralino}) = (1500,100)$ GeV, respectively.

Based on background estimates derived from the MC samples 
described in section~\ref{sec:samples}, the 
selection requirements were optimized as a function
of \MET, \MEFF and \ptg by maximizing the expected discovery sensitivity of the analysis,
for each of the two signal benchmark points.
The selected
values of the minimum requirements on all three
optimization parameters were found to be very similar
for the low and high neutralino-mass benchmark points,
leading to the definition of a single signal region (SR).
The selection requirements for this SR
are shown in Table~\ref{tab:SR}.

\begin{table}[bp]
    \centering
    \caption{
        Requirements defining the signal region (SR)
        and the $W\gamma\gamma$ CR referred to in Section~\ref{sec:backgrounds}.
    }
    \begin{tabular}{l l}
    \hline \hline
    SR                                  & $W\gamma\gamma$ CR\\
    \hline
    2 tight photons with $\pt > 75$~GeV & 2 tight photons with $\pt > 50$~GeV   \\
                                        & 1 $e$ or $\mu$ with $\pt > 25$~GeV    \\
    $\dphijm > 0.5$                     & $\dphijm > 0.5$   \\
    $\MET  > 175$~GeV                   & $50 < \MET < 175$~GeV  \\
    $\MEFF > 1500$~GeV                  & $N(\mathrm{jets}) < 3$ \\
                                        & $m_{e \gamma} \notin 83\text{--}97$~GeV \\
    \hline\hline
    \end{tabular}
    \label{tab:SR}
\end{table}

%-------------------------------------------------------------------------------
\section{Background estimation}
\label{sec:backgrounds}

Processes that contribute to the Standard Model background of diphoton final states
can be divided into three primary components.
The largest contribution to the inclusive diphoton spectrum is the
``QCD background,'' which can be further 
divided into a contribution from two real photons produced in association with jets, and a
``jet-faking-photon'' contribution arising from
\mbox{$\gamma$+jet} and multijet events for which one or both reconstructed photons
are faked by a jet, typically by producing a $\pi^{0}\rightarrow\gamma\gamma$ decay
that is misidentified as a prompt photon.
An ``electron-faking-photon background'' arises predominantly from $W$, $Z$, and $\; t\bar{t}$ events,
possibly accompanied by additional jets and/or photons, for which an electron is misidentified as a photon.
Electron-to-photon misidentification is due primarily to instances for which an
electron radiates a high-momentum photon as it traverses the material of the ATLAS inner detector.
Last, an ``irreducible background'' arises from $W\gamma\gamma$ and $Z\gamma\gamma$ events.
These backgrounds are estimated with a combination of data-driven and simulation-based methods
described as follows.

The component of the QCD background arising from real diphoton events ($\gamma\gamma$)
is estimated directly from diphoton MC events, 
reweighted as function of \MET
and the number of selected jets to match the respective distributions for the 
inclusive diphoton sample in the range $\MET <$ \unit[100]{GeV}.
While this background dominates the inclusive diphoton sample,
it is very steeply falling in \MET, making it small relative 
to backgrounds with real $\MET$ for
$\MET \gtrsim$ \unit[100]{GeV}, independent of the reweighting.

The component of the QCD background arising from jets faking photons and the 
background arising from electrons faking photons are
both estimated with a data-driven ``fake-factor'' method, for which events in data
samples enriched in the background of interest are weighted by factors parameterizing
the misidentification rate.

To estimate the jet-faking-photon fake-factor, the jet-faking-photon background is enriched by using an inverted isolation
requirement, selecting events only if they contain one or more non-isolated photons.
The relative probability of an energy cluster being reconstructed as an isolated,
rather than non-isolated, 
photon is known as the photon-isolation fake factor, and is measured in an
orthogonal ``non-tight'' sample of photons. The selection of this sample
requires that all the tight photon identification requirements be satisfied,
with the exception that at least one of the requirements on the calorimeter variables
defined only with the first (strip) layer of the electromagnetic calorimeter
fails. This leads to a sample enriched in identified (non-tight) photons
that are actually $\pi^0$s within jets.
The correlation between the isolation variable and the photon identification
requirements was found to be small and to have no significant impact
on the estimation of the jet-faking-photon fake-factor.
The fake factors depend upon $\pt$ and $\eta$, and vary between 10\% and 30\%.
The jet-faking photon background is then estimated by weighting
events with non-isolated photons by the applicable photon-isolation fake factor.

The electron-faking-photon background is estimated with a similar fake-factor method.
For this case, the electron-faking-photon background is enriched by selecting events with
a reconstructed electron instead of a second photon.
Fake factors for electrons being misidentified as photons are then measured
by comparing the ratio of reconstructed $e\gamma$ to $ee$ events arising from $Z$ bosons decaying to 
electron--positron pairs,
selected within the mass range of 75--105\:GeV.
The electron-faking-photon background is then estimated by weighting selected
$e\gamma$ events by their corresponding fake factors, which are typically a few percent.

The irreducible background from $W\gamma\gamma$ events is estimated with MC simulation;
however, because it is a potentially dominant background contribution, 
the overall normalization is derived in a $\ell\gamma\gamma$ control region ($W\gamma\gamma$ CR) as follows.
Events in the $W\gamma\gamma$ CR are required to have two tight, isolated photons with $\pt > 50$ GeV,
and exactly one selected lepton (electron or muon) with $\pt > 25$ GeV.
As with the SR, events are required to have $\dphijm > 0.5$, so that the direction of
the missing transverse momentum vector is not aligned with that of any high-\pt jet.
To ensure that the control sample has no overlap with the signal
region, events are discarded if $\MET > 175$ GeV. While these requirements target
$W\gamma\gamma$ production, they also are expected to select appreciable 
backgrounds from $t\bar{t}\gamma$, $Z\gamma$ and $Z\gamma\gamma$ events, and thus
additional requirements are applied. To suppress $t\bar{t}\gamma$ contributions to the $W\gamma\gamma$ CR,
events are discarded if they contain more than two selected jets.
To suppress $Z\gamma$ contributions, events are discarded if there is
an $e$--$\gamma$ pair in the events with $83 < m_{e\gamma} < 97$ GeV.
Finally, to  
suppress $Z\gamma\gamma$ contributions, events with $\MET <50$~GeV are discarded.
The event selection requirements for the $W\gamma\gamma$ CR are summarized in Table~\ref{tab:SR}.
A total of seven events are observed in this $W\gamma\gamma$ control region,
of which 1.6 are expected to arise from sources other than $W\gamma\gamma$
production. The MC expectation for the $W\gamma\gamma$ process is 1.9 events,
leading to a $W\gamma\gamma$ scale factor of $2.9 \pm 1.4$, assuming that no GGM signal events contaminate 
the $W\gamma\gamma$ CR.
This scale factor is consistent with that of 
the corresponding $\sqrt{s} =$ \unit[8]{TeV} analysis~\cite{ref:photon_X_2012}, and is reconciled
by a large and uncertain NLO correction to the
$W\gamma\gamma$ production cross section
that depends strongly upon the momentum of the $W\gamma\gamma$ system~\cite{Bozzi:2011wwa}.
When setting limits on specific signal models,
a simultaneous fit to the control region and the signal region
is performed, allowing both the signal and $W\gamma\gamma$ contributions
to float to their best-fit values. 

Last, the irreducible background from $Z(\to\nu\nu)\gamma\gamma$ 
events, the only background without a data-derived normalization,
is estimated with simulation and found to be 0.02 events.
A $\pm 100\%$ uncertainty is conservatively applied to account for
modelling uncertainties~\cite{STDM-2014-01}.

A summary of the background contributions to the signal region
is presented in Table~\ref{tab:total_background}.
The QCD background can be traced to a few hundredths of an event at
high \MET and high \MEFF, but no events are observed for either
the diphoton Monte Carlo or the jet-faking-photon control sample when the full signal region requirements are applied.
Relaxing the \MEFF requirement, and
using a conservative extrapolation of the expected QCD background as a function of \MEFF,
the combined QCD background is
estimated to be $0.05^{+0.20}_{-0.05}$ events.
The estimate of the electron-faking-photon background is established by the
presence of two $e\gamma$ events in the background model passing the SR requirements,
yielding a background estimate of $0.03 \pm 0.02$ events
after application of the fake-factor weights.
Summing all background
contributions, a total of $0.27^{+0.22}_{-0.10}$
SM events are expected in the SR,
with the largest contribution, $0.17 \pm 0.08$ events,
expected to arise from $W\gamma\gamma$ production.
The background modelling was found to agree well in several validation regions,
including the inclusive high-\pt{} diphoton sample,
as well as event selections with relaxed \MEFF{} and \MET{} requirements relative to those of the SR.

\begin{table}[bp]
\centering
\caption{
    Summary of background estimates by source, and total combined background,
    in the signal region.
    The uncertainties shown include the total statistical and systematic uncertainty.
    Also shown is the 
    %are the observed number of events in the SR (`Data') and the 
    expected number of signal events for the benchmark points
    $(m_{\gluino},m_{\neutralino}) = (1500,100)$ and $(m_{\gluino},m_{\neutralino}) = (1500,1300)$,
    where all masses are in GeV.
}
\begin{tabular}{lr}
\hline \hline
Source           & Number of events  \\
\vspace{-4 mm} \\
\hline   \\
\vspace{-9 mm} \\
QCD ($\gamma\gamma$, $\gamma$j, jj)     &    $0.05^{+0.20}_{-0.05}$ \\
$e\rightarrow\gamma$ fakes              &           $0.03 \pm 0.02$ \\
$W\gamma\gamma$                         &           $0.17 \pm 0.08$ \\
$Z\gamma\gamma$                         &           $0.02 \pm 0.02$ \\
\vspace{-4 mm} \\
\hline   \\
\vspace{-9 mm} \\
Sum                                     &  $0.27^{+ 0.22}_{- 0.10}$ \\
\vspace{-4 mm} \\
\hline
$(m_{\gluino},m_{\neutralino}) = (1500,100)$  &                       7.0 \\
$(m_{\gluino},m_{\neutralino}) = (1500,1300)$ &                       8.0 \\
\hline \hline
\end{tabular}
\label{tab:total_background}
\end{table}

%-------------------------------------------------------------------------------
\section{Signal efficiencies and uncertainties}
\label{sec:signal_eff}
%-------------------------------------------------------------------------------

GGM signal acceptances and efficiencies are estimated using MC simulation for each simulated
point in the gluino--bino parameter space,
and vary significantly across this space.
For example, for a gluino mass of \unit[1600]{GeV}, the 
acceptance-times-efficiency product varies between 14\% and 28\%, reaching a minimum
as the NLSP mass approaches the $Z$ boson mass, below which the photonic branching
fraction of the NLSP rises to unity.
Table~\ref{tab:sys:ggm} summarizes the contributions to the systematic uncertainty
of the signal acceptance-times-efficiency, which are discussed below.

Making use of a bootstrap method~\cite{ATLAS-CONF-2012-048},
the efficiency of the diphoton trigger is determined
to be greater than 99\%, with an uncertainty of less than 1\%.
The uncertainty in the integrated luminosity is $\pm 2.1\%$. 
It is derived, following a methodology similar to that detailed in Ref.~\cite{Aad:2013ucp},
from a calibration of the luminosity scale using x-y beam-separation scans performed in August 2015.
%Based on the methodology of Ref.~\cite{Aad:2013ucp}, and using
%$x$--$y$ beam-separation scans performed in August 2015,
%the uncertainty in the integrated luminosity is estimated to be $\pm 2.1\%$.

The reconstruction and identification efficiency for tight, isolated photons is estimated with
complementary data-driven methods~\cite{Aaboud:2016yuq}.
Photons selected kinematically as originating from radiative decays of a $Z$ boson
($Z \rightarrow \ell^+ \ell^- \gamma$ events)
are used to study the photon reconstruction efficiency as a function of $\pt$ and $\eta$.
Independent measurements making use of a tag-and-probe approach with
$Z \to ee$ events, with one of the electrons
used to probe the calorimeter response to electromagnetic depositions, also provide
information about the photon reconstruction efficiency. For photons with $\pt > 75$ GeV,
the identification efficiency varies between 93\% and 99\%, depending
on the values of the photon $\pt$ and $|\eta|$ and whether the photon converted in the inner detector.
The uncertainty also depends upon these factors, 
and is generally no more than a few percent. 

Uncertainties in
the photon and jet energy scales lead to uncertainties in
the signal acceptance-times-efficiency that vary across
the GGM parameter space, and can contribute the dominant source of
acceptance-times-efficiency uncertainty in certain
regions of the parameter space.
The photon energy scale is
determined using samples of $Z \to ee$ and $J/\psi \to ee$ events~\cite{PERF-2013-05}.
The jet-energy scale uncertainty is constrained
from an assessment of the effect of uncertainties in the modelling of jet properties
and by varying the response to differing jet flavour composition
in MC simulations, as well as from \emph{in situ} measurements with 
8 TeV dijet data~\cite{ATL-PHYS-PUB-2015-015, ATLAS-CONF-2015-037}.

Uncertainties in the values of whole-event observables, such as \MET and \MEFF, arise from
uncertainties in the energy of the underlying objects from which they are constructed.
Uncertainties in the \MET{} soft term contribute negligibly,
as varying the soft term to reflect systematic
uncertainties was found to
affect the signal acceptance-times-efficiency
by less than 0.1\%.
The uncertainty due to pile-up is estimated by varying the mean of the distribution of the number
of interactions per bunch crossing overlaid in the simulation by $\pm 11$\%.

Including the contribution from the statistical limitations
of the MC samples used to model the GGM parameter space,
the quadrature sum of the individual 
systematic uncertainties in the signal reconstruction
efficiency is, on average, about $4$\%.
Adding the uncertainty in the integrated luminosity gives a total systematic uncertainty of about 5\%.

\begin{table}[bp]
    \centering
    \caption{
      Summary of individual and total contributions to the systematic uncertainty of the signal  
      acceptance-times-efficiency. Relative uncertainties are shown in percent, and as the 
      average over the full range of the ($m_{\gluino}$,$m_{\neutralino}$) grid. Because the 
      individual contributions are averaged over the grid only for that particular source, 
      the average total uncertainty is not exactly equal to the quadrature sum of the 
      individual average uncertainties.
    }
    \begin{tabular}{ l r }
      \hline \hline
      Source of systematic uncertainty         & Value\\
      \hline
      Luminosity                    & \unit[2.1]{\%} \\
      Photon identification         & \unit[3.0]{\%}  \\
      Photon energy scale           & \unit[0.2]{\%}  \\
      Photon energy resolution      & \unit[0.2]{\%}  \\
      Jet energy scale              & \unit[0.4]{\%}  \\
      Jet energy resolution         & \unit[0.3]{\%}  \\
      \met{} soft term              & $< 0.1$\%   \\
      Pile-up uncertainty           & \unit[$1.8$]{\%} \\
      MC statistics                 & \unit[$2.3$]{\%} \\
      \hline
      Total experimental uncert.    &  \unit[$4.7$]{\%} \\
      \hline \hline
    \end{tabular}
    \label{tab:sys:ggm}
\end{table}

%-------------------------------------------------------------------------------
\section{Results}
\label{sec:result}
%-------------------------------------------------------------------------------

An accounting of the numbers of events observed in the SR
after the successive application of the selection
requirements is shown in Table~\ref{tab:cutflow}
along with the size of the expected SM
background. After the full selection is applied,
no events are observed in the SR, to be compared 
to an expectation of $0.27^{+ 0.22}_{-0.10}$ SM events.

\begin{table}[bp]
    \centering
    \caption{
    Numbers of events observed in the SR
    after the successive application of the selection
    requirements, as well as the size of the expected SM
    background.
    }
    \begin{tabular}{lr}
    \hline \hline
    Requirement     &  Number of Events         \\
    \hline
    Two photons, $\ptg > 75$ &  4982     \\
    $\dphijm > 0.5$     &    4724 \\
    $\MEFF > 1500$~GeV  &   1    \\
    $\MET  > 175$~GeV   &   0   \\
\vspace{-4 mm} \\
\hline   \\
\vspace{-9 mm} \\
Expeected SM background                                     &  $0.27^{+ 0.22}_{- 0.10}$ \\
\vspace{-4 mm} \\
%\hline
%Data                                    &                         0 \\
\hline \hline
    \end{tabular}
    \label{tab:cutflow}
\end{table}

Based on the observation of zero events in the SR and the
magnitude of the estimated SM background expectation and uncertainty, an
upper limit is set on the number of events
from any scenario of physics beyond the SM, using the
profile likelihood and $CL_s$ prescriptions~\cite{Read:2002hq}.
The various sources of experimental uncertainty, including those
in the background expectation, are treated as
Gaussian-distributed nuisance parameters in the likelihood definition.
Assuming that no events due to physical processes beyond those
of the SM populate the $\ell\gamma\gamma$ CR used to estimate \Wgg background,
the observed 95\% confidence-level (CL) upper limit on the number of non-SM events 
in the SR is found to be 3.0.
Taking into account the
integrated luminosity of \integLumi,
this numbers-of-event limit translates into a 95\% CL upper
limit on the visible cross section for new physics, defined by the
product of cross section, branching fraction, acceptance and
efficiency, of 0.93 fb.

By considering, in addition, the value and uncertainty of the
acceptance-times-efficiency of the selection requirements associated with
the SR,
as well as the NLO (+NLL) GGM
cross section~\cite{Beenakker:1996ch,Kulesza:2008jb,Kulesza:2009kq,Beenakker:2009ha,Beenakker:2011fu},
which varies steeply with gluino mass,
95\% CL lower limits may be set on the mass of the gluino as a function of
the mass of the lighter bino-like neutralino, in the
context of the GGM scenario described in Section~\ref{sec:ggm}.

\begin{figure}[tp]
    \centering
    \includegraphics[width=0.95\textwidth]{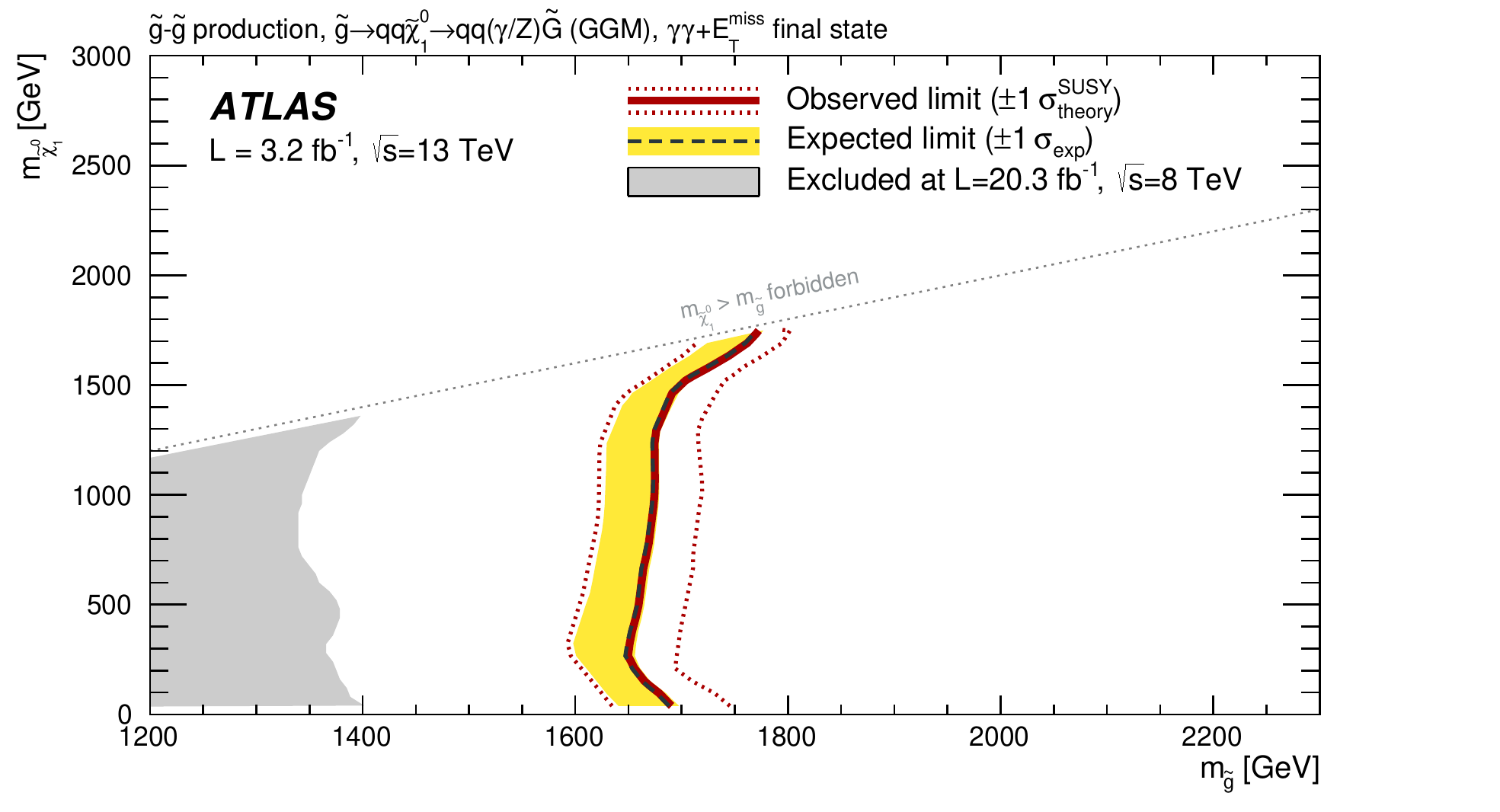}
    \caption{
      Exclusion limits in the neutralino--gluino mass plane at 95\% CL.
      The observed limits are exhibited for the nominal SUSY model cross section,
      as well as for a SUSY cross section increased and lowered
      by one standard deviation of the cross-section systematic uncertainty.
      Also shown is the expected limit, as well as the $\pm 1$ standard-deviation
      range of the expected limit, which is asymmetric because of the low count expected. 
      Because the background expectation is close to zero
      and the observed number of events is zero, the expected and observed limits nearly overlap.
      The previous limit from ATLAS using 8 TeV data~\cite{ref:photon_X_2012} is shown in grey.
    }
    \label{fig:di_gluino_limits}
\end{figure}

The resulting observed limit on the gluino mass
is exhibited, as a function
of neutralino mass, in
Figure~\ref{fig:di_gluino_limits}.
For the purpose of establishing these model-dependent limits,
the \Wgg background estimate
and the limit on the possible number of events from new physics
are extracted from a simultaneous fit to the SR and \Wgg control region,
although for a gluino mass in the range of the observed limit
the signal contamination in the \Wgg control sample is less
than 0.03 events for any value of the neutralino mass.
Also shown for this figure
is the expected limit, including its
statistical and background uncertainty range,
as well as observed limits for a SUSY model cross section $\pm 1$ standard deviation of theoretical
uncertainty from its central value.
Because the background expectation is close
to zero and no events are observed in data, the expected and observed limits nearly overlap.
The observed lower limit on the gluino mass is observed to be roughly independent
of neutralino mass, reaching a minimum value of approximately \unit[1650]{GeV}
at a neutralino mass of \unit[250]{GeV}.

%-------------------------------------------------------------------------------
\section{Conclusion}
\label{sec:conclusion}
%-------------------------------------------------------------------------------

A search has been made for a diphoton + \MET final state using the ATLAS
detector at the Large Hadron Collider in \integLumi
of proton--proton collision data taken at a
centre-of-mass energy of \unit[13]{TeV} in 2015.
At least two photon candidates with $\pt>\unit[75]{GeV}$ are required,
as well as minimum values of \unit[175]{GeV} and \unit[1500]{GeV}
of the missing transverse momentum and effective mass of the event, respectively.
The resulting signal region targets 
events with pair-produced high-mass
gluinos each decaying to either a
high-mass or low-mass bino-like neutralino.
Using a combination of data-driven and direct Monte Carlo approaches, 
the SM background is estimated to be $0.27^{+0.22}_{-0.10}$ events, 
with most of the expected background arising from
the production of a $W$ boson in association with two energetic photons. 
No events are observed in the signal region;
considering the expected background and its uncertainty,
this observation implies model-independent 95\% CL upper limits of 3.0 events (0.93 fb) on
the number of events (visible cross section) due to physics beyond the
Standard Model. In the context of a generalized model of gauge-mediated
supersymmetry breaking with a bino-like NLSP, this leads to a lower limit of \unit[1650]{GeV}
on the mass of a degenerate octet of gluino states, independent of the 
mass of the lighter bino-like neutralino. This extends the corresponding
limit of \unit[1340]{GeV}
derived from a similar
analysis of \unit[8]{TeV} data by the ATLAS Collaboration.

%-------------------------------------------------------------------------------
\section*{Acknowledgements}
\label{sec:acknowledgements}
%-------------------------------------------------------------------------------

% Acknowledgements for papers with collision data
% Version 15-Apr-2016

% Standard acknowledgements start here
%----------------------------------------------
We thank CERN for the very successful operation of the LHC, as well as the
support staff from our institutions without whom ATLAS could not be
operated efficiently.

We acknowledge the support of ANPCyT, Argentina; YerPhI, Armenia; ARC, Australia; BMWFW and FWF, Austria; ANAS, Azerbaijan; SSTC, Belarus; CNPq and FAPESP, Brazil; NSERC, NRC and CFI, Canada; CERN; CONICYT, Chile; CAS, MOST and NSFC, China; COLCIENCIAS, Colombia; MSMT CR, MPO CR and VSC CR, Czech Republic; DNRF and DNSRC, Denmark; IN2P3-CNRS, CEA-DSM/IRFU, France; GNSF, Georgia; BMBF, HGF, and MPG, Germany; GSRT, Greece; RGC, Hong Kong SAR, China; ISF, I-CORE and Benoziyo Center, Israel; INFN, Italy; MEXT and JSPS, Japan; CNRST, Morocco; FOM and NWO, Netherlands; RCN, Norway; MNiSW and NCN, Poland; FCT, Portugal; MNE/IFA, Romania; MES of Russia and NRC KI, Russian Federation; JINR; MESTD, Serbia; MSSR, Slovakia; ARRS and MIZ\v{S}, Slovenia; DST/NRF, South Africa; MINECO, Spain; SRC and Wallenberg Foundation, Sweden; SERI, SNSF and Cantons of Bern and Geneva, Switzerland; MOST, Taiwan; TAEK, Turkey; STFC, United Kingdom; DOE and NSF, United States of America. In addition, individual groups and members have received support from BCKDF, the Canada Council, CANARIE, CRC, Compute Canada, FQRNT, and the Ontario Innovation Trust, Canada; EPLANET, ERC, FP7, Horizon 2020 and Marie Sk{\l}odowska-Curie Actions, European Union; Investissements d'Avenir Labex and Idex, ANR, R{\'e}gion Auvergne and Fondation Partager le Savoir, France; DFG and AvH Foundation, Germany; Herakleitos, Thales and Aristeia programmes co-financed by EU-ESF and the Greek NSRF; BSF, GIF and Minerva, Israel; BRF, Norway; Generalitat de Catalunya, Generalitat Valenciana, Spain; the Royal Society and Leverhulme Trust, United Kingdom.

The crucial computing support from all WLCG partners is acknowledged
gratefully, in particular from CERN and the ATLAS Tier-1 facilities at
TRIUMF (Canada), NDGF (Denmark, Norway, Sweden), CC-IN2P3 (France),
KIT/GridKA (Germany), INFN-CNAF (Italy), NL-T1 (Netherlands), PIC (Spain),
ASGC (Taiwan), RAL (UK) and BNL (USA) and in the Tier-2 facilities
worldwide.
%----------------------------------------------

%-------------------------------------------------------------------------------
\clearpage
%\appendix
%\part*{Appendix}
%\addcontentsline{toc}{part}{Appendix}
%-------------------------------------------------------------------------------
%\input{tex/appendix}

%-------------------------------------------------------------------------------
% If you use biblatex and either biber or bibtex to process the bibliography
% just say \printbibliography here
\printbibliography
% If you want to use the traditional BibTeX you need to use the syntax below.
%\bibliographystyle{bibtex/bst/atlasBibStyleWoTitle}
%\bibliography{SUSYPhotons2015,bibtex/bib/ATLAS}
%-------------------------------------------------------------------------------

%-------------------------------------------------------------------------------
% Print the list of contributors to the analysis
% The argument gives the fraction of the text width used for the names
%-------------------------------------------------------------------------------
%\clearpage
%\PrintAtlasContribute{0.30}

%-------------------------------------------------------------------------------
\clearpage
%\appendix
%\part*{Auxiliary material}
%\addcontentsline{toc}{part}{Auxiliary material}
%\input{tex/aux_material}
%-------------------------------------------------------------------------------

% Authorlist
\newpage
% ATLAS Collaboration author list
% Data extracted on 23-Jun-2016 for paper reference SUSY-2016-04
%\documentclass[11pt]{article}
%\usepackage{a4wide}%\begin{document}
\begin{flushleft}
{\Large The ATLAS Collaboration}

\bigskip

M.~Aaboud$^\textrm{\scriptsize 135d}$,
G.~Aad$^\textrm{\scriptsize 86}$,
B.~Abbott$^\textrm{\scriptsize 113}$,
J.~Abdallah$^\textrm{\scriptsize 64}$,
O.~Abdinov$^\textrm{\scriptsize 12}$,
B.~Abeloos$^\textrm{\scriptsize 117}$,
R.~Aben$^\textrm{\scriptsize 107}$,
O.S.~AbouZeid$^\textrm{\scriptsize 137}$,
N.L.~Abraham$^\textrm{\scriptsize 149}$,
H.~Abramowicz$^\textrm{\scriptsize 153}$
% ...
\bigskip
\\
% ...
$^{12}$ Institute of Physics, Azerbaijan Academy of Sciences, Baku, Azerbaijan\\
% ...
$^{64}$ University of Iowa, Iowa City IA, United States of America\\
% ...
$^{86}$ CPPM, Aix-Marseille Universit{\'e} and CNRS/IN2P3, Marseille, France\\
% ...
$^{107}$ Nikhef National Institute for Subatomic Physics and University of Amsterdam, Amsterdam, Netherlands\\
% ...
$^{113}$ Homer L. Dodge Department of Physics and Astronomy, University of Oklahoma, Norman OK, United States of America\\
% ...
$^{117}$ LAL, Univ. Paris-Sud, CNRS/IN2P3, Universit{\'e} Paris-Saclay, Orsay, France\\
% ...
$^{135}$ $^{(a)}$ Facult{\'e} des Sciences Ain Chock, R{\'e}seau Universitaire de Physique des Hautes Energies - Universit{\'e} Hassan II, Casablanca; $^{(b)}$ Centre National de l'Energie des Sciences Techniques Nucleaires, Rabat; $^{(c)}$ Facult{\'e} des Sciences Semlalia, Universit{\'e} Cadi Ayyad, LPHEA-Marrakech; $^{(d)}$ Facult{\'e} des Sciences, Universit{\'e} Mohamed Premier and LPTPM, Oujda; $^{(e)}$ Facult{\'e} des sciences, Universit{\'e} Mohammed V, Rabat, Morocco\\
% ...
$^{137}$ Santa Cruz Institute for Particle Physics, University of California Santa Cruz, Santa Cruz CA, United States of America\\
% ...
$^{149}$ Department of Physics and Astronomy, University of Sussex, Brighton, United Kingdom\\
% ...
$^{153}$ Raymond and Beverly Sackler School of Physics and Astronomy, Tel Aviv University, Tel Aviv, Israel\\
% ...
$^{a}$ Also at Department of Physics, King's College London, London, United Kingdom\\
$^{b}$ Also at Institute of Physics, Azerbaijan Academy of Sciences, Baku, Azerbaijan\\
$^{c}$ Also at Novosibirsk State University, Novosibirsk, Russia\\
$^{d}$ Also at TRIUMF, Vancouver BC, Canada\\
$^{e}$ Also at Department of Physics {\&} Astronomy, University of Louisville, Louisville, KY, United States of America\\
% ...
\end{flushleft}

%\end{document}
% Created with xml2latex.py

\end{document}